\documentclass[10pt,aps,prc,superscriptaddress,preprintnumbers,showpacs,nofootinbib,endfloats,showkeys,onecolumn]{revtex4}
\usepackage[dvips]{graphics,graphicx}
\usepackage{amsmath}
\usepackage{wrapft}
\begin{document}
\preprint{WU-HEP-03-5}
\title{Transverse momentum dependence of Hanbury Brown-Twiss radii of
pions from a perfectly opaque source with hydrodynamic flow}
\author{Kenji Morita}
\email{morita@hep.phys.waseda.ac.jp}
\affiliation{Department of Physics, Waseda University, Tokyo 169-8555, Japan}
\author{Shin Muroya}
\email{muroya@yukawa.kyoto-u.ac.jp}
\affiliation{Tokuyama Women's College, Shunan, Yamaguchi 745-8511,
Japan}

\begin{abstract}
 We investigate the transverse momentum dependence of pion HBT radii 
on the basis of 
 a hydrodynamical model. Recent experimental data show
that $R_{\text{out}}/R_{\text{side}} < 1$, which suggests a strong opaqueness
 of the source. In addition to the opaqueness naturally caused by 
 transverse flow, we introduce an extrinsic opacity by imposing
 restrictions on the pion emission angle. Comparing the HBT radii obtained
 from the normal Cooper-Frye prescription and the opaque emission
 prescription, we find that $R_{\text{out}}/R_{\text{side}}$ 
 is less than unity only for small values of the transverse momentum 
with an opaque
 source. However, HBT radii for large values of the transverse momentum 
are dominated by
 the transverse flow effect and are  affected less by the source
 opaqueness.
\end{abstract}

\maketitle
\section{Introduction}\label{sec:intro}

Pion interferometry is one of the most promising tools for use 
in high energy
heavy ion collision experiments designed to explore the states of matter
under extreme conditions. As is well known, the Hanbury Brown-Twiss (HBT)
effect, caused by the symmetry of the wave function of identical bosons,
provides information concerning the 
geometry of the source via two-particle intensity
correlation functions \cite{Weiner_PREP}. Recently, the
Relativistic Heavy Ion Collider (RHIC) at Brookhaven National Laboratory
(BNL) has started to operate at extremely high energies. This opens a new
frontier of heavy ion collision experiments. Experimental data obtained 
at the partial collision energy $\sqrt{s}=130A$ GeV 
and preliminary data at $\sqrt{s}=200A$ GeV have been 
reported \cite{QM2002}. One of the most interesting but puzzling
results is that derived from the pion HBT data \cite{STAR_HBT}.
Usually, the outward HBT
radius $R_{\text{out}}$ is believed to be
larger than the sideward HBT radius 
$R_{\text{side}}$ by an amount equal to 
the duration time, i.e.  $R_{\text{out}}^2 \sim R_{\text{side}}^2 +
\beta_{\bot}^2
 (\Delta t)^2 $. 
A hydrodynamical model with a phase transition naively predicts large
outward HBT radii around the RHIC energy because of the prolonged lifetime
of the fluid due to the existence of the phase
transition \cite{Hydro_Longtau,Rischke_NPA608}.  The obtained HBT radii
in the RHIC experiments, however, reveal the surprising feature that
$R_{\text{out}}$
is almost the same as (or even smaller than) $R_{\text{side}}$. 
The ratio $R_{\text{out}}/R_{\text{side}}$, which is proposed 
for a good indicator of the long emission 
duration \cite{Rischke_NPA608},  exhibits a
slight decrease around unity with the pair transverse momentum. This
strange result is sometimes called the ``HBT puzzle''.

Let us briefly survey the present situation regarding HBT radii.
The collision process is governed by strong and multi-body
interactions, including multiparticle production. 
At this time, we are far from the dynamical
description of the entire process in terms of 
the fundamental theory.
Though pions possess information only at their
freeze-out, because of the strong interaction, the two-pion
correlation function provides the space-time distribution of the
freeze-out point and the history of the space-time evolution, which is
subject to the equation of state. Therefore, a dynamical model is
indispensable for understanding the
HBT data, and a hydrodynamical approach is quite suitable for this
purpose. In addition, recent experimental data concerning
 an anisotropic flow
($v_2$) obtained in the mid-rapidity region strongly 
support the validity of the
hydrodynamical picture at the energies typically studied with the 
RHIC \cite{Kolb_PLB500}.
However, conventional hydrodynamical model analyses that reproduce
single-particle spectra (and elliptic flow in Ref.~\cite{Heinz_NPA702})
result in unsatisfactory HBT radii \cite{Heinz_NPA702,Morita_PRCRapid}.
Space-time evolution with a smooth crossover transition equation of
state provides a small improvement, but the resulting predictions for 
the HBT radii are 
still far from agreeing with the data \cite{Zschiesche_PRC65}. Though
 smaller HBT radii can be obtained, especially in the longitudinal
direction, by introducing the chemical freeze-out in addition to 
the thermal
freeze-out \cite{Hirano_PRC66} (see also Ref.~\cite{Kolb_PRC67}),
this decrease is still insufficient to account 
for the experimental results.
Some transport calculations indicate that the 
creation of dense partonic matter has a tendency to improve the
HBT radii \cite{Lin_PRL89}.  
The problem will be the description of the hadronic stage
and the subsequent freeze-out.
A possible solution of this problem may be to introduce 
a modification of the freeze-out hypersurface in 
 hydrodynamical models, on which the number of the 
emitted particle is usually carried out by 
using the Cooper-Frye prescription
\cite{Cooper_Frye} with a sharp three-dimensional hypersurface.
A more sophisticated treatment of the freeze-out combined with the
hadronic transport calculation is still not sufficient to 
obtain predictions consistent with 
experiment \cite{Soff_PRL}.

Considering the
meaning of the HBT radii given by second order moments of the emission
function \cite{Chapman_PRL74}, the experimentally obtained results 
that $R_{\text{out}} \sim R_{\text{side}}$ indicate an ``opaque
source'' \cite{Heiselberg_EPJC1,Tomasik_nucl9805016}, from which particles
are emitted only on a thin surface. 
It has been shown that transverse flow in the hydrodynamic evolution also
naturally causes opaque features of the HBT radii \cite{Morita_PRC},
and such effects are already
automatically taken into account in the calculation. However, 
this effect is
not sufficient to realize consistency with the 
experimental data. Therefore, we must introduce 
additional mechanisms that increase the source opacity.\footnote{In
Ref.~\cite{McLerran_hep0205028}, HBT radii are investigated with an
opaque source model inferred from quark and gluon evaporation. 
Their analysis given there, however, is based on (1+1)-dimensional 
longitudinal
expansion and ignores transverse flow.}
In the present paper, we investigate the HBT radii on the bases of
a hydrodynamical model \cite{Morita_PRCRapid}. By putting a
restriction on the emission angle of pions, we introduce  complete
opaqueness of the source given by the hydrodynamical model. Comparing
the HBT radii obtained from normal emission with those from this opaque
source model, we study the transverse momentum dependence of the HBT
radii and clarify its origin. 

In the next section, we  briefly
explain our model. In $\S$\ref{c2}, a description of the opaque source
model is given. Section \ref{sec:result} is devoted to  results
and discussion.

\section{Model}\label{sec:model}

We describe the space-time evolution of hot matter created in
$\sqrt{s}= 200A$ GeV Au+Au central collisions at the RHIC using the
hydrodynamical model presented in Ref.~\cite{Morita_PRCRapid}. We can
fit single-particle distributions of charged hadrons, such as the 
pseudorapidity distribution from the PHOBOS
collaboration \cite{PHOBOS_dndeta200}, the 
net-proton rapidity distribution
from the BRAHMS
collaboration \cite{BRAHMS_netproton}, and the identified
transverse momentum distributions from the PHENIX collaboration
\cite{PHENIX_pt}, by adjusting parameters in the initial matter
distributions. Here, we adopt the following functional form of the initial net
baryon number distribution:
\begin{gather}
 n_{\text{B}}(\tau_0,\eta,r)=n_{\text{B0}}
  \left\{
   \exp\left[
	-\frac{(|\eta|-\eta_{\text{D}})^2}{2\sigma_{\text{D}}^2}
      \right]\theta(|\eta|-\eta_0) \nonumber
   +\exp\left[
       -\frac{(\eta_0-\eta_{\text{D}})^2}{2\sigma_{\text{D}}^2}
       \right]\theta(\eta_0-|\eta|)\right\} \\
 \times \exp\left[-\frac{(r-r_0)^2}{2\sigma_r^2}\theta(r-r_0)\right].
\end{gather}
This form is employed in order to reproduce the 
\textit{flat} net-proton distribution for
$|y|~\leq~1$ obtained by the BRAHMS collaboration \cite{BRAHMS_netproton},
\ The parameter set is listed in Table.~\ref{tbl:parameter}. The particle distributions are calculated using 
the Cooper-Frye prescription \cite{Cooper_Frye}.\footnote{In the present
calculation, we include not only the time-like component of $d\sigma_\mu$, as
done in Ref.~\cite{Morita_PRCRapid}, but also the 
space-like component of surface
elements.}  We also take into account the resonance 
decay contribution, as in
Ref.~\cite{Morita_PRCRapid}. Note that our calculation does
not reproduce absolute numbers of kaons and (anti-)protons, because we
assume the single freeze-out condition. 
Incorporating chemical freeze-out
\cite{Hirano_PRC66,Teaney_nucl0204023,Kolb_PRC67}, or an additional
assumption for the initial stage that leads to stronger transverse flow
\cite{Eskola_hep0206230}, makes it possible to reproduce both the 
$p_{\text{T}}$
slope and the yield of each particle species. But such modifications
do not affect our main argument below.

\begin{table}
 \begin{center}
  \caption{\label{tbl:parameter}Parameter set for 200$A$ GeV Au+Au
  collisions at RHIC.}
  \vspace*{-0.4cm}
  \begin{tabular}[t]{cccccccc}\hline\hline
   $\varepsilon_{\text{max}}$ & $n_{\text{B0}}$ & $\eta_0$ & $\sigma_r$
   & $\sigma_\eta$ & $\eta_D$ & $\sigma_D$ & $T_{\text{f}}$  \\ \hline
   6.9 GeV/fm$^3$ & 0.22 fm$^{-3}$ & 1.3 & 1.0 fm & 1.4 & 2.6 & 0.92
   & 130 MeV \\ \hline
  \end{tabular}
  \end{center}
\end{table}

\section{Two-pion correlation function from an opaque source}\label{c2}

The two-pion intensity correlation function for a chaotic source is
given by
\cite{Shuryak_PLB44}
\begin{equation}
 C_2(q,K)=1+\frac{|I(q,K)|^2}{I(0,k_1)I(0,k_2)}\label{eq:c2}
\end{equation}
where $k_1$ and $k_2$ are the \textit{on-shell} four-momenta of the 
two emitted
pions, and $q$ and $K$ are the relative and average 
four-momentum defined as
$q=k_1-k_2$ and $K=(k_1+k_2)/2$, respectively. We choose $I(q,K)$ as
\begin{equation}
 I(q,K)= \int_\Sigma K\cdot d\sigma (x) e^{iq\cdot x} f(x,K),\label{eq:iqx}
\end{equation}
where $f(x,K)$ is the Bose-Einstein distribution 
function $[\exp(K\cdot U(x)
/T(x))-1]^{-1}$, so that $I(0,k)$ reduces to the 
Cooper-Frye formula of a
single-particle distribution \cite{Chapman_PLB340}.
Note that $K$ is the \textit{off-shell} four-momentum,
 and $\Sigma$ denotes
that the integration is carried out over the
 3-dimensional freeze-out hypersurface determined by
$T=T_{\text{f}}$.

\begin{figure}[ht]
 \includegraphics{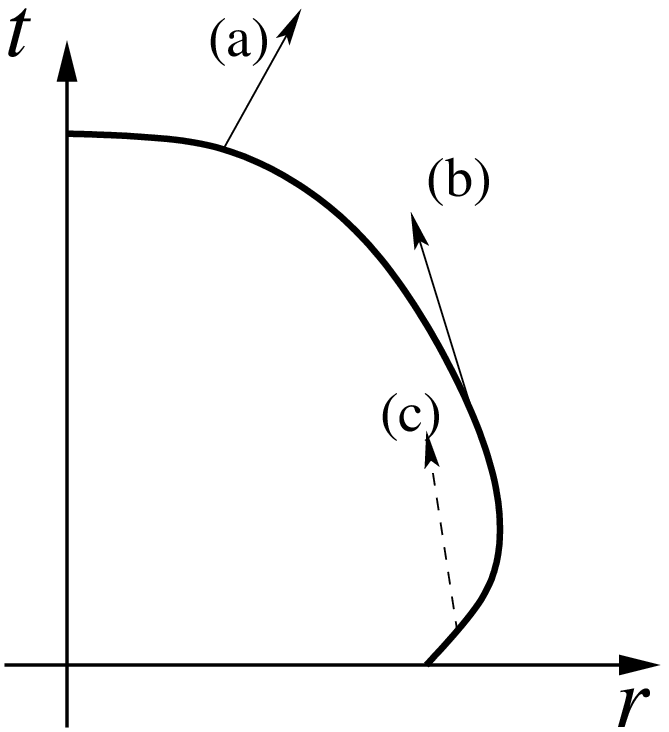}
 \caption{\label{fig:emission}A schematic view of the 
 freeze-out
 hypersurface (thick curve) in the $r$-$t$ plane ($\eta=0$) 
 and the directions
 (arrows) of emitted particles. (a) and (b) are allowed directions,
 but (c) is forbidden.}
\end{figure}

As mentioned in $\S$\ref{sec:intro}, the opaqueness of the source 
represents a
possible solution of the HBT puzzle. Here, we introduce a
simple opaque source model by imposing two conditions on the emission from
the freeze-out hypersurface.

The first condition is the space-time geometrical constraint that
a particle can not be emitted into the fluid; i.e.,
``inward'' emission is forbidden. The assumption of local
thermal equilibrium implies isotropic particle emission in the local
rest frame of a fluid element. 
Therefore, inward emission naturally occurs even after 
a Lorentz boost to the center-of-mass system, though its
contribution to the multiplicity is known to be small. 
In the Cooper-Frye freeze-out prescription, 
the number of particles absorbed from time-like surfaces can be 
considered as
negative \cite{Sinyukov_ZPHYS43,Bernard_NPA605,Bugaev_NPA606,Anderlik_PRC59}.
We omit such emissions by introducing the step function
$\theta(p\cdot d\sigma)$ for $p=k_1$ and $k_2$ in the surface
integrations. (See also Fig.~\ref{fig:emission}.) 

\begin{figure}[tb]
 \centerline{\includegraphics[width=3.375in]{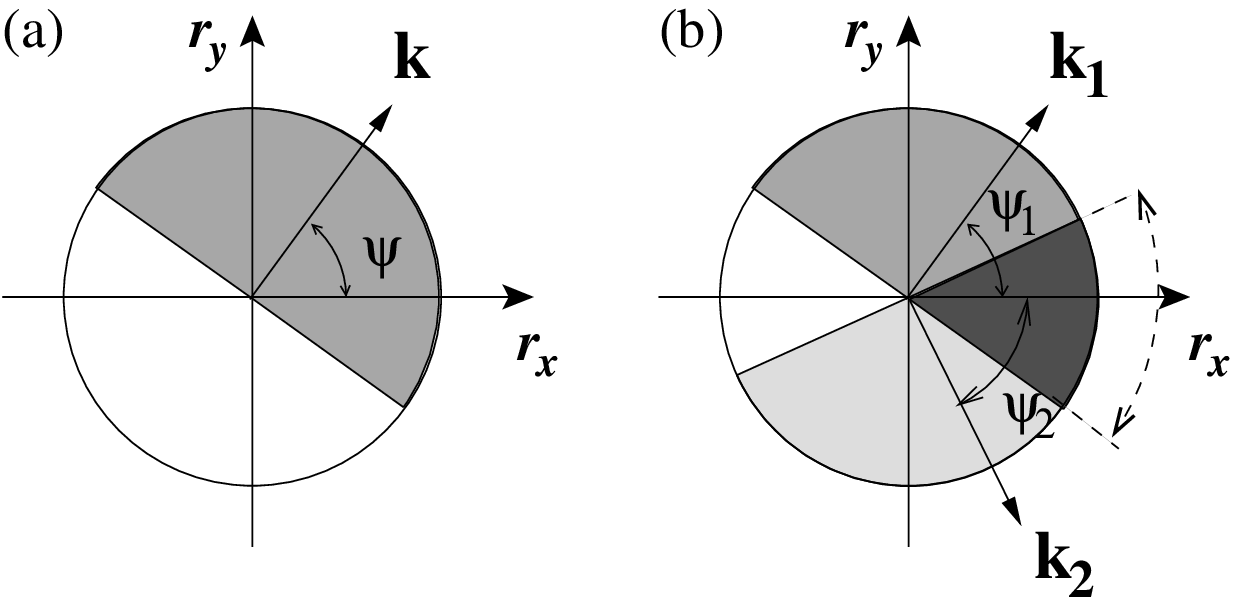}}
 \caption{\label{fig:xyplane}Emission angle constraint in $x$-$y$ plane.
 In (a), the filled region represents the 
allowed region that satisfies the
 constraint. Figure (b) describes the two-particle case, 
where the overlapping area indicated by the darkest shading is that 
in which the constraint is satisfied for both
 particles.}
\end{figure}

The second condition is a purely three-dimensional geometrical constraint
that is characterized by the factor
$\theta(\cos(\phi-\psi))$,\footnote{This condition can also be expressed as
$\theta(\boldsymbol{k_{\text{T}}\cdot r_{\text{T}}})$, where
$\boldsymbol{r_{\text{T}}}$ is the radial vector in the
transverse plane.} where
$\phi = \tan^{-1}(r_y / r_x)$ is the
azimuthal angle of the emission point, $\psi =\tan^{-1}(k_y/k_x)$ is
the azimuthal emission angle of the 
 particle (Fig.~\ref{fig:xyplane}), and $\theta$ is the step
function. This constraint prohibits the 
emissions in the direction opposite to the radial flow velocity, 
which occur naturally by virtue of the assumption of the local thermal
equilibrium. 
For example, for a particle with $ k_x =k_{\text{T}} $ and $k_y=0$, 
we allow emissions from space-time points at $r_x > 0$,
which corresponds to the limited range of azimuthal
angles of emission points $-\pi/2 \leq \phi \leq \pi /2$ .
Hence, we regard the source constructed with these two conditions as a
perfectly opaque source, from which particles 
cannot be emitted from the side opposite 
 to the detector. 

Note that the above
constraints should be imposed on each particle, while the two-particle
correlation function can be expressed as the Fourier transform of the 
(pseudo-) single-particle distribution function. (See also
Fig.~\ref{fig:xyplane}(b).) Therefore, we input the
constraints on each emitted particle into both numerator and denominator
of Eq.~\eqref{eq:c2} as
\begin{equation}
 C_2(q,K)
  =1+
  \frac{\displaystyle \left|\int_\Sigma K\cdot d\sigma f(x,K) e^{iq\cdot x} 
	  \Theta(x,k_1)\Theta(x,k_2)\right|^2  }
  {\displaystyle \int_\Sigma k_1\cdot d\sigma f(x,k_1) \Theta(x,k_1) 
  \int_\Sigma k_2\cdot d\sigma' f(x',k_2) \Theta(x',k_2)}\label{eq:opaquec2}
\end{equation}
where $\Theta(x,k_i) \equiv \theta(k_i\cdot
d\sigma)\theta[\cos(\phi-\psi_i)]$ and $\psi_i =
\tan^{-1}(k_{iy}/k_{ix})$ for $i=1,2$. Under these constraints, energy
conservation between the fluid and emitted particles through the
freeze-out process is violated. We have estimated the loss of energy 
in a system with
the conditions described above to be 11\%. 
This means that 89\% of emitted
energy quanta satisfy the conditions. The resultant multiplicity
$dN/d\eta_p$ at $\eta_p =0$, where $\eta_p$ denotes the pseudorapidity
$\eta_p = 0.5\ln[(|\boldsymbol{p}|+p_z)/(|\boldsymbol{p}|-p_z)]$,
decreases from 618 to 593. This loss can be considered 
negligible for the discussion below. 
Some attempts have been made to avoid this kind of 
violation by improving the freeze-out prescription
\cite{Anderlik_PRC59}, but it would be a formidable 
task to incorporate both the constraints and the conservation 
low in a self-consistent manner, 
and this is beyond the scope of this paper. 

\section{Results and discussions}\label{sec:result}

The HBT radii are obtained through a 3-dimensional $\chi^2$-fit
to the correlation functions \eqref{eq:c2} with the form
\begin{equation}
 C_{\text{2fit}}(q,K)
=1+\exp[-R_{\text{side}}^2(K) q_{\text{side}}^2
-R_{\text{out}}^2(K) q_{\text{out}}^2
-R_{\text{long}}^2(K) q_{\text{long}}^2] \label{eq:c2fit} .
\end{equation}

\begin{figure}[b]
 \centerline{\includegraphics[width=6cm]{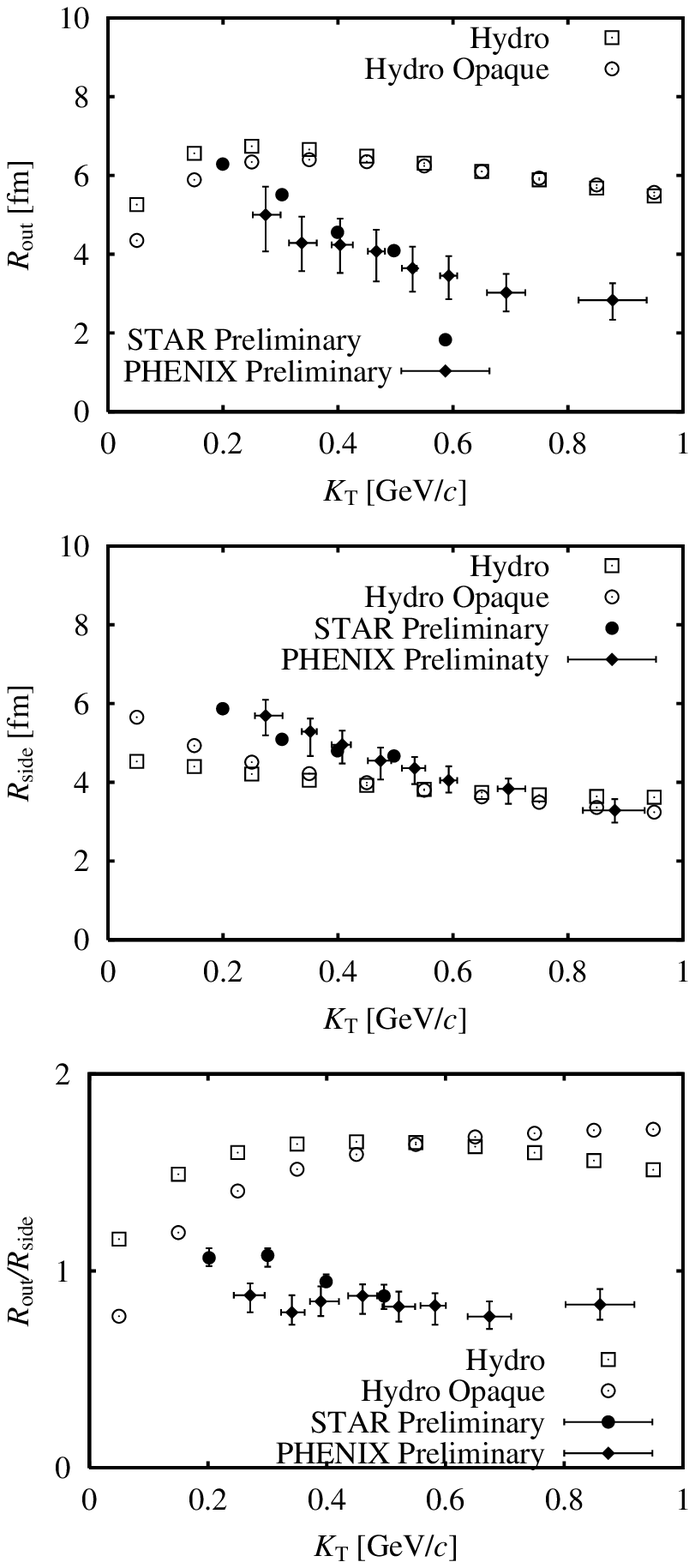}}
 \caption{\label{fig:radii}HBT radii for 200$A$ GeV Au+Au
 collisions as functions of the averaged 
 transverse momentum $K_{\text{T}}$. 
 From top to bottom, the open squares represent the results from our
 model with normal freeze-out, given by Eqs.~\eqref{eq:c2} and
 \eqref{eq:iqx}. The results from opaque freeze-out
 [Eq.~\eqref{eq:opaquec2}] are represented by the open circles. The
 solid circles and squares represent the STAR and PHENIX experimental
 results, respectively.}
\end{figure}

In this section, we compare the ``normal'' HBT radii calculated using
Eqs.~\eqref{eq:c2} and \eqref{eq:iqx} and the ``opaque'' case calculated
using Eq.~\eqref{eq:opaquec2}. We focus on $R_{\text{out}}$ and
$R_{\text{side}}$, 
because the purpose of this paper is to 
clarify the effects of the transverse
dynamics, i.e., the flow effect and opacity effect. 
$R_{\text{long}}$
does not reflect such effects, because its $K_{\text{T}}$ dependence
originates mainly from rapid longitudinal expansion
\cite{Hama_PRD37}.
In the numerical evaluation of the correlation function, the experimental
window effect was
simply ignored for the better understanding of the opacity effect. 
The following approximate expressions for the HBT radii 
in terms of second-order moments of the source function are convenient
\cite{Chapman_PRC}:
\begin{align}
 R_{\text{side}}^2(\boldsymbol{k}) &= \langle \tilde{r_y}^2 \rangle, \\
 R_{\text{out}}^2(\boldsymbol{k}) &= \langle (\tilde{r_x}-\beta_{\perp}\tilde{t})^2 \rangle \nonumber \\
 &= \langle \tilde{r_x}^2 \rangle -2\beta_{\perp}
 \langle \tilde{r_x}\tilde{t} \rangle 
+\beta_{\perp}^2 \langle \tilde{t}^2 \rangle. \label{eq:rout}
\end{align}
 Here,
\begin{equation}
 \langle A(x) \rangle (\boldsymbol{k}) \equiv 
  \frac{\displaystyle \int_\Sigma k\cdot d\sigma f(x,k) A(x)}
  {\displaystyle \int_\Sigma k\cdot d\sigma f(x,k)},
\end{equation}
where $\tilde{x}\equiv x-\langle x \rangle $ and
$\beta_{\perp}=k_{\text{T}}/E_{\boldsymbol{k}}$.
These are good approximation for both of the emission prescriptions.

Figure \ref{fig:radii} displays the results for HBT radii together with
recent preliminary experimental results from the STAR \cite{STAR_HBT200}
and the PHENIX \cite{PHENIX_HBT200} obtained from $\sqrt{s}=200A$ GeV
Au+Au collisions.

\begin{figure}[htb]
 \centerline{\includegraphics[width=3.375in]{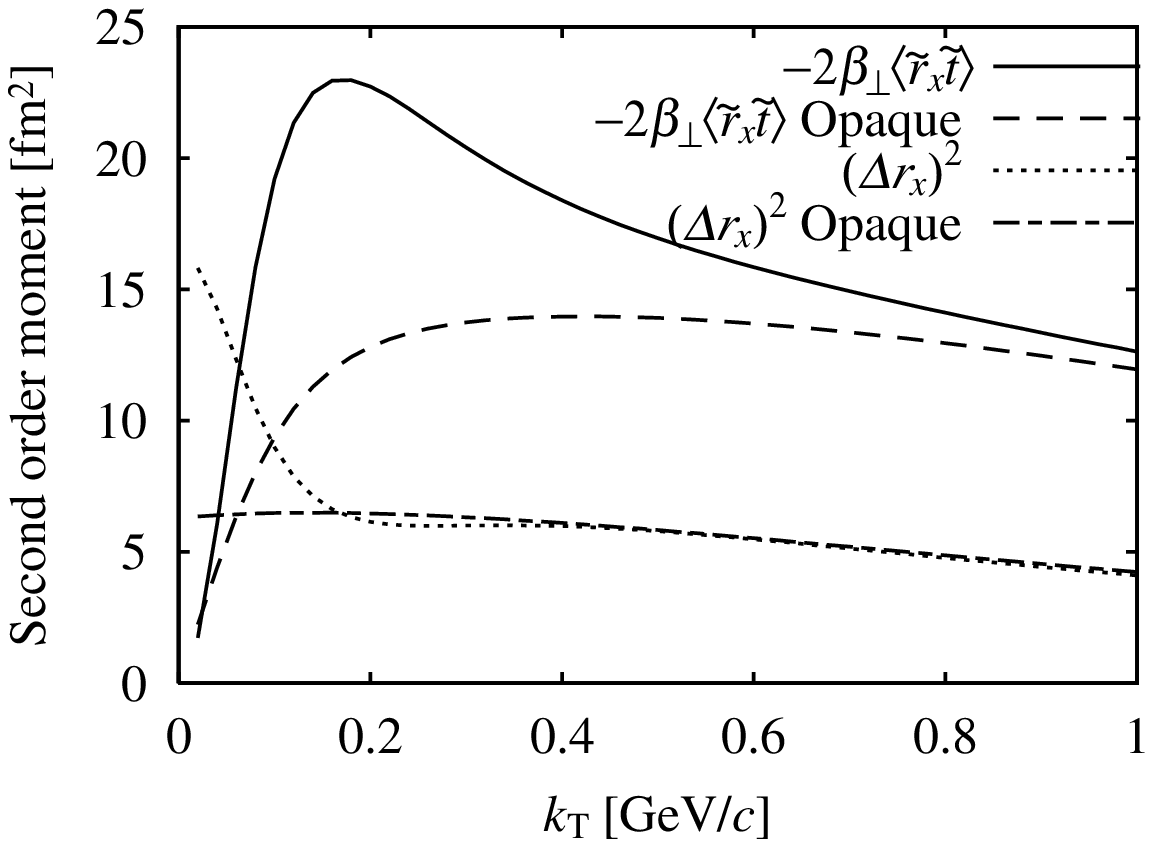}}
 \caption{\label{fig:moment}$r_x$-$t$ correlations and source width in the
 outward ($x$) direction. }
\end{figure}
\begin{figure}[htb]
 \centerline{\includegraphics[width=3.375in]{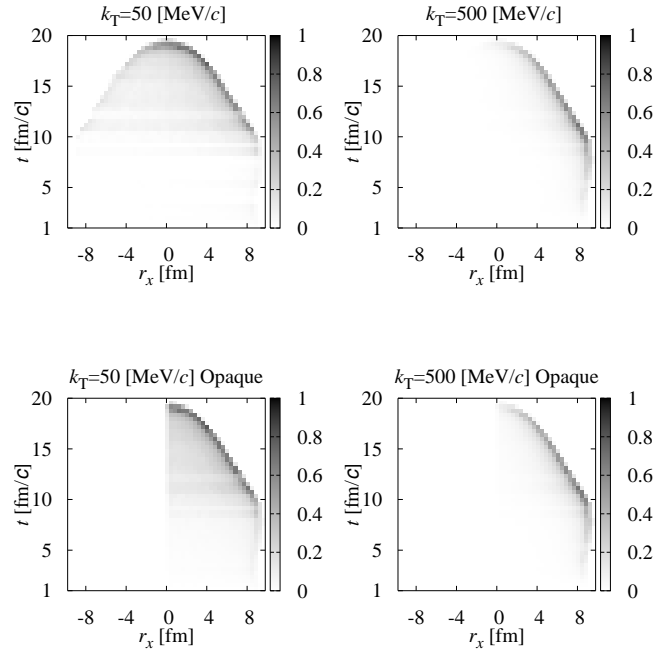}}
 \caption{\label{fig:sxt}Source function in the $r_x$-$t$ plane. The upper
 graphs are for the case of the normal freeze-out procedure, and the 
 lower ones are for the opaque emission case. For each freeze-out, the
 small transverse momentum case ($k_{\text{T}}$ = 50 MeV/$c$, left row)
 and large transverse momentum case ($k_{\text{T}}$ = 500 MeV/$c$, right
 row) are presented.}
\end{figure}
\begin{figure}[b]
 \centerline{\includegraphics[width=3.375in]{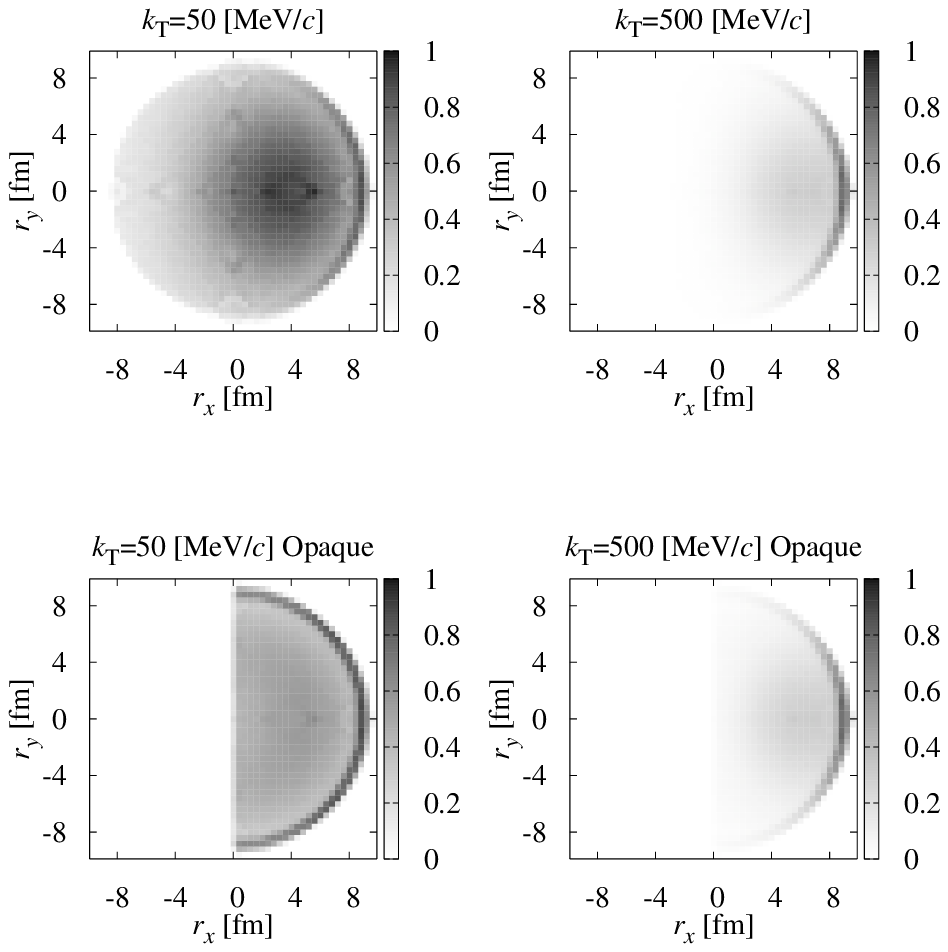}}
 \caption{\label{fig:sxy}Source function in the $r_x$-$r_y$ plane. 
 As in the previous figure, the normal and opaque emission cases are 
 considered for $k_{\text{T}}$ = 50 MeV/$c$ and 500 MeV/$c$, respectively.}
\end{figure}

The results of the normal freeze-out prescription exhibit 
deviations similar to those seen in 
the 130$A$ GeV results \cite{Morita_PRCRapid}. In particular,
$R_{\text{side}}$ is smaller and decreases less steeply 
with increasing $K_{\text{T}}$, $R_{\text{out}}$ is larger, 
and $R_{\text{out}}/R_{\text{side}}$ is much larger, which 
corresponds to a long emission
duration due to the phase transition. Despite the strong restriction on
the emission direction by Eq.~\eqref{eq:opaquec2}, only a slight
improvement of $R_{\text{side}}$ and $R_{\text{out}}$ is seen.
In the opaque source model, $R_{\text{side}}$ becomes larger at small
$K_{\text{T}}$, and its $K_{\text{T}}$ dependence becomes slightly
steeper than in the normal emission case. $R_{\text{out}}$ becomes
slightly 
smaller at small $K_{\text{T}}$, though no improvement can be seen beyond
$K_{\text{T}}\sim 0.5$
GeV/$c$. Consequently, the ratio of $R_{\text{out}}$ to $R_{\text{side}}$ 
improves to a value less than unity in the smallest
$K_{\text{T}}$ bin. Nevertheless, in the larger $K_{\text{T}}$ region,
the opaque emission does not yield results consistent with 
 the experimental data.
These results suggest that the emission angle restriction does not
affect the correlation function at larger $K_{\text{T}}$. After all,
the flow effect dominates in this region. 
This fact can be intuitively understood by calculating $r_x$-$t$
correlations and $\langle \tilde{r_x}^2 \rangle$ (Fig.~\ref{fig:moment})
and by plotting the source functions in the $r_x$-$t$ (Fig.~\ref{fig:sxt})
and $r_x$-$r_y$ (Fig.~\ref{fig:sxy}) planes. The source functions are
calculated as
\begin{align}
 S_{xt}(r_x,t;k_{\text{T}})&= * \int_{|\eta|\leq 1} k\cdot d\sigma(x') 
 f(x',k) \delta(r_x-r_x')\delta(t-t'), \\
 S_{xy}(r_x,r_y;k_{\text{T}})&= * \int k \cdot d\sigma(x') f(x',k)
 \delta(r_x-r_x')\delta(r_y-r_y').
\end{align}
Each function is calculated at mid-rapidity and is normalized 
such that its maximum
value is unity. For $S_{xt}(r_x,t;k_{\text{T}})$, integration 
over the space-time rapidity $\eta\equiv
0.5\ln[(t+z)/(t-z)]$ is carried out with 
$|\eta|\leq 1$ in order to obtain
a clear emission probability distribution. For the opaque emission model,
$\Theta(x,k)$ is inserted into the above definitions as in
Eq.~\eqref{eq:opaquec2}. 

\begin{figure}[b]
 \centerline{\includegraphics[width=3.375in]{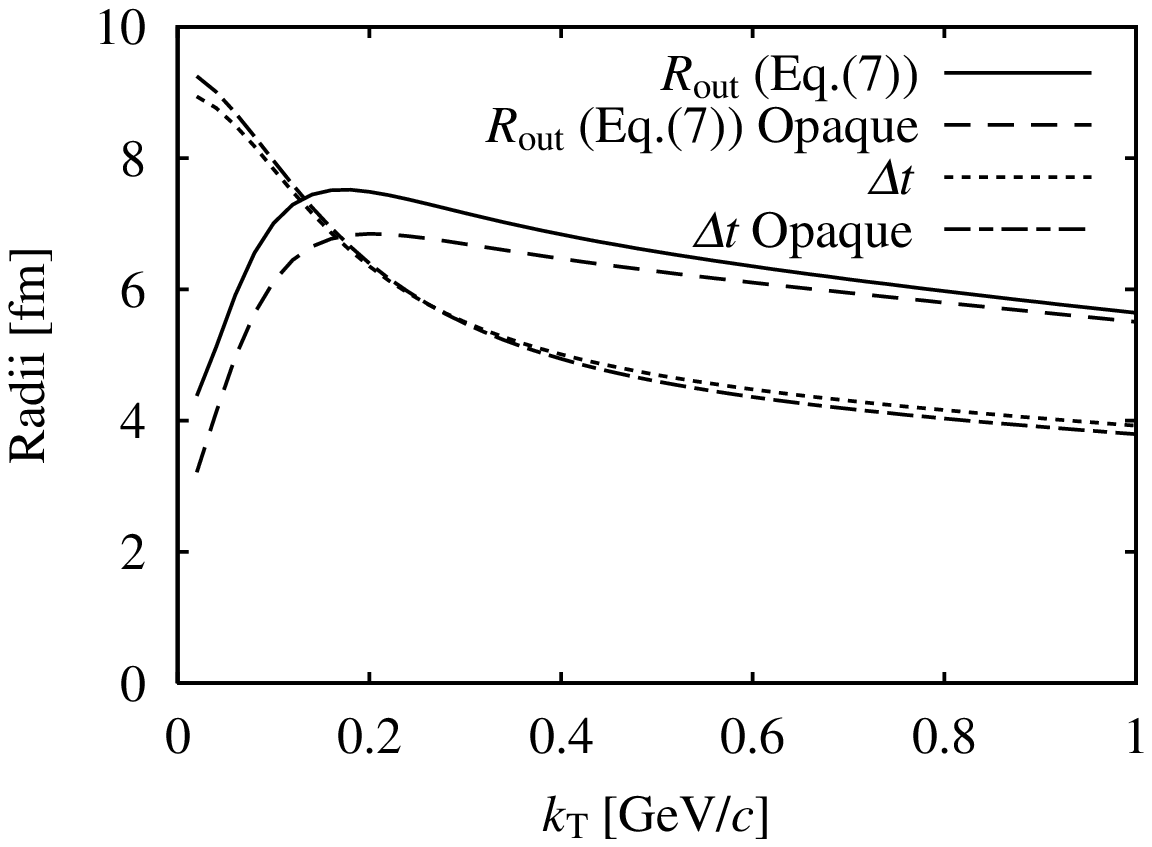}}
 \caption{\label{fig:deltat}$R_{\text{out}}$ calculated from
 Eq.~\eqref{eq:rout} and the time duration, $\Delta t$.}
\end{figure}

Figure~\ref{fig:moment} reveals a large decrease of 
$\Delta r_x = \sqrt{\langle \tilde{r_x}^2\rangle}$ at small
$k_{\text{T}}$ in the opaque emission model, as expected. 
This behavior is
clearly caused by the emission angle restriction factor
$\theta(\cos(\phi-\psi_i))$, which cuts off emission from $x < 0$ for
$k_{\text{T}}=(k_{\text{T}},0)$, as seen from the left column
($k_{\text{T}}=50$ MeV/$c$) of Fig.~\ref{fig:sxy}.  However, such a distinct
difference disappears at higher $k_{\text{T}}$, because the flow causes a
strong suppression and enhancement through the Boltzmann factor (right
column). Thus, as $k_{\text{T}}$ increases, the relative effect of opacity
 becomes weaker in the presence of transverse flow. Another factor,
$-2\beta_\bot \langle \tilde{r_x}\tilde{t} \rangle$, in  $R_{\text{out}}$
exhibits strong dependence on $k_{\text{T}}$, as seen in Fig.~\ref{fig:moment},
though it does not affect $R_{\text{out}}$ greatly. In the
$k_{\text{T}}$ region satisfying
$(\Delta r_x)_{\text{opaque}} \ll (\Delta r_x)_{\text{normal}}$, 
the prefactor $\beta_{\bot}$ is so small that the second term of
Eq.~\eqref{eq:rout} gives only a small contribution to $R_{\text{out}}$.
The decrease of $\langle \tilde{r_x}\tilde{t} \rangle$ 
in the opaque emission model is due to the larger $\langle r_x
\rangle$, which is a natural consequence of emission only from 
the region $r_x > 0$.
The leading contribution in this region, however, is $\Delta t$,
which is 6 fm$/c$ for $k_{\text{T}}=200$ MeV/$c$
(Fig.~\ref{fig:deltat}). Therefore, 
the reduction of $\langle \tilde{r_x}\tilde{t} \rangle$ does not reduce
$R_{\text{out}}$ so much. From Fig.~\ref{fig:sxy}, some increase of the 
emissivity at the edge can be seen due to the cutoff of time-like
surface emission. (Also, as seen from Fig.~\ref{fig:emission}, surface elements
at the edge naturally have a time-like part.) This fact results in the
slight increase of $R_{\text{side}}$ for small $k_{\text{T}}$, because
$R_{\text{side}}$ can be considered the width of the source along the $y$
direction.

In the present paper, we have demonstrated that a naive opaque emission
model in which we forbid emissions through dense media does not account
for the ``HBT puzzle'' if collective transverse flow exists. 
The opaqueness caused by the dense matter preventing the pions from
passing through such media affects the HBT radii only at small transverse
momentum. As a result, smaller values of $R_{\text{out}}$ 
than $R_{\text{side}}$ are 
obtained only for small $k_{\text{T}}$. Incorporating the opacity effect
transforms the shape of the source function. Nevertheless,
transverse flow dominates the source function for large $k_{\text{T}}$;
 modification of the source function by the opacity is so slight that
the $k_{\text{T}}$ dependence of the HBT radii is still dominated by the
transverse flow effect.

As mentioned above, the present study constitutes a trial 
studying the 
modification of the source shape. One can consider other
possibilities. For example, a viscosity correction 
\cite{Teaney_QM02} and $\rho$
meson broadening \cite{Soff_QM02} have been examined. It also should be noted
that a thermal model analysis yields acceptable agreement
\cite{Broniowski_HBT}. This agreement results from the fact that
 their model used in that study has a positive $r_x$-$t$ correlation
due to the choice of the freeze-out hypersurface. However, that surface
was simply added by hand, and is not the result of a dynamical
calculation. Further investigation is required to solve
 the puzzle.\footnote{Recently it has been suggested that a partial
Coulomb correction can improve the HBT radii \cite{Enokizono_JPS}.}

\section*{Acknowledgements}
The authors would like to thank Drs. S.~Dat\'{e}, T.~Hatsuda, T.~Matsui,
 A.~Nakamura, H.~Nakamura, H.~Nakazato and I.~Ohba for fruitful
 discussions and comments. They are also indebted to Prof. T.~Sugitate
 for the treatment of the Ref.~\cite{Enokizono_JPS}. This work is
 partially supported by the
 Ministry of Education, Culture, Sports, Science and Technology, 
Japan (Grant No.~13135221)
 and a Waseda University Grant for Special Research 
Projects (No.~2003A-095).

\end{document}